\documentclass{JAC2003}

\usepackage{graphicx}
\usepackage{booktabs}

\newcommand{\SV}{\sigma_\mathrm{VBAND}}
\newcommand{\SZED}{\sigma_\mathrm{ZED}}
\newcommand{\SVLN}{\sigma_\mathrm{VLN}}
\newcommand{\sqrts}{\sqrt{s}}
\newcommand{\sqrtsNN}{\sqrt{s_\mathrm{NN}}}

\newcommand{\pp}{\mbox{pp}}
\newcommand{\PbPb}{\mbox{Pb--Pb}}
\newcommand{\pPb}{\mbox{p--Pb}}
\newcommand{\Pbp}{\mbox{Pb--p}}
\newcommand{\pA}{\mbox{p--A}}
\newcommand{\Ap}{\mbox{A--p}}
\renewcommand{\AA}{\mbox{A--A}}

\begin{document}

\title{Reference cross section measurements with ALICE\\ in \pp\ and \PbPb\ collisions at LHC}

\author{K. Oyama, University of Heidelberg, Germany\\
        for the ALICE Collaboration }

\maketitle

\begin{abstract}
   Cross sections of reference trigger processes were obtained based on beam property measurements in dedicated
   luminosity calibration experiments (van der Meer scans).
   These cross-sections are essential for absolute cross section determinations of physics processes.
   The reference cross sections are presented for \pp\ collisions at two energies; $\sqrts=2.76$~TeV and 7~TeV,
   and for \PbPb\ collisions at $\sqrtsNN=2.76$~TeV
   together with a discussion of the systematic uncertainty 
   originating from beam intensity and rate measurement uncertainties.
\end{abstract}

\section{INTRODUCTION}

The determination of the cross section of a reference trigger process ($\sigma_{\mathrm{trig}}$)
enables the calculation of the luminosity $\mathcal{L}$ using
the relation $\mathcal{L} = R_{\mathrm{trig}} / \sigma_{\mathrm{trig}}$ where $R_{\mathrm{trig}}$ is the rate of the reference trigger,
and, hence, the measurement of absolute cross sections of physics processes.

In the ALICE experiment\cite{ALICE}, such reference cross sections have been measured using
the van der Meer (vdM) scan method\cite{VDM}.
$R_{\mathrm{trig}}$ is measured as a function of the beam separation and provides information
on the spatial convolution (effective beam sizes) of the two colliding beams.

The uncertainty on the reference cross-section is part of the systematic
uncertainty of all subsequent cross section measurements.
Hence, a trigger setup that is stable in time and that provides a high rate is selected for vdM scans
in order to obtain the longest possible validity and to obtain a statistical error below the
systematic uncertainties discussed below. 
Therefore the necessary precision for the reference cross section is given by the analysis
that requires the highest precision, and it is different for different collision systems: proton-proton (\pp),
proton-nucleus (\pA), and nucleus-nucleus (\AA) collisions.

\subsection{Precision required in \pp}

In heavy-ion collision experiments, particle production is often compared
to the extrapolation from elementary \pp\ collisions via binary scaling.
The nuclear modification factor $R_{\mathrm{AA}}^{(X)}$ for a given process $X$ is defined as the ratio between
the yield in \AA\ collisions $N_{\mathrm{AA}}^{(X)}/N_{\mathrm{evt}}$ and the yield expected by scaling
the \pp\ cross section $\sigma_{\mathrm{pp}}^{(X)}$ by the average nuclear overlap function
$\langle T_{\mathrm{AA}} \rangle$:
\begin{equation}
 R_{\mathrm{AA}}^{(X)}
 = \frac{N_{\mathrm{AA}}^{(X)}/N_{\mathrm{evt}}}{\langle T_{\mathrm{AA}} \rangle \cdot \sigma_{\mathrm{pp}}^{(X)}}.
\end{equation}
In order to quantify the strength of nuclear effects, the desired precision of $R_{\mathrm{AA}}^{(X)}$ is typically $<$10\%.
Thus a precision of the order of 5\% on $\sigma_{\mathrm{pp}}^{(X)}$ is required for it to not dominate
the overall uncertainty.

On the other hand, ALICE measures \pp\ collisions not only as reference for \AA\ but
also as a field of study in its own right.

Measurements such as the total inelastic cross section\cite{CSPAPER} or the $J/\psi$
cross section\cite{JPSI} can achieve precisions below 3\%,
and therefore a 1-2\% precision for the reference cross sections is required 
in order to avoid dominance of the total uncertainty.

\subsection{Precision required in \pA}

Similarly to \pp\ measurements, the main purpose of \pA\ data analysis is to obtain a reference for \AA\ analysis.
In order to avoid that the  uncertainty on the \pA\ reference cross section dominates the reference for \AA,
a 2-3\% precision is desired.

One particular physics measurement in \pA, the gluon distribution in protons (nuclei) using ultra-peripheral
$J/\psi$ production in \pA\ (\Ap) collisions, requires rather good precision.
In the \pPb\ and \Pbp\ runs planned in 2013, the statistical uncertainties are expected to be
$\sim$6\% and $\sim$3\%, respectively.
The systematic uncertainty from the measurement itself is at the 5\% level.
Thus a 3\% precision for the reference cross section will be sufficient.

\subsection{Precision required in \AA}

Generally in \AA\ collisions, such as \PbPb, most physics analyses do not require the direct measurement of cross sections
but rely on particle yields per interaction using the Glauber model fit approach\cite{CENTRALITY,GLAUBER}.
Here the \pp\ inelastic cross section is more important as described above.
However, there are a few exceptional cases requiring the direct measurements of cross sections in \AA\ collisions.

Probing the gluon distribution function in nuclei is one of the important measurements to be performed by
looking at $J/\psi$ production in ultra-peripheral \PbPb\ collisions.
At the LHC energy, ALICE can explore the regime at Bjorken's $\mathrm{x}$ close to 10$^{-3}$,
and can constrain the shadowing effect.
Systematic uncertainties on the measurement will be $>$5\%, thus a 2-3\% error on the luminosity is sufficient.

The electromagnetic nuclear dissociation (EMD) cross section represents another important measurement.
The analysis requires a cross section measurement directly using the vdM scan in \AA.
The model uncertainty to be compared to the experimental error is about 5\% for the single EMD process
and 2\% for the low multiplicity neutron emission process\cite{RELDIS,ALICEEMD}.
Here the luminosity uncertainty becomes crucial as other sources of uncertainties are relatively small (3-4\%).
Thus a 1-2\% uncertainty on the reference cross section is satisfactory.

\section{Detector Setup}

\begin{figure}[t]
    \centering
    \includegraphics*[width=80mm]{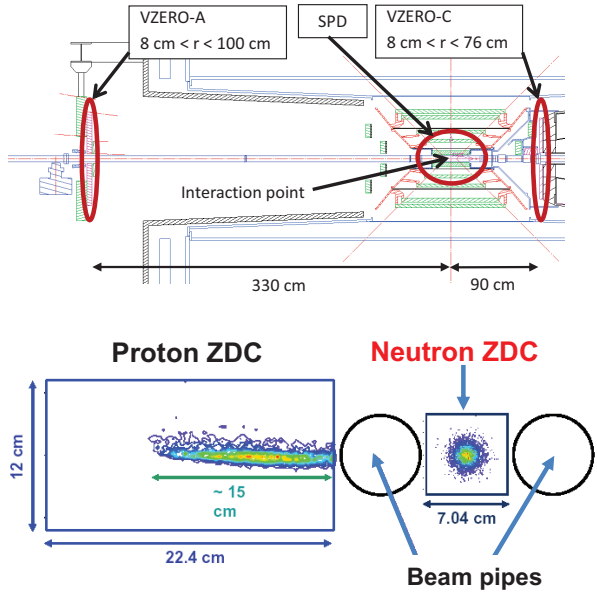}
    \caption{
    Schematic view showing the approximate positions and shapes of the ALICE VZERO detectors (top) and ZDC detectors (bottom).
    Top panel: side view of the beam pipe at the ALICE IP.
    The locations of the VZERO arrays are indicated.
    Bottom panel: cross section view (perpendicular to the beam line), of the ZDC system.
    }
    \label{fig-V0ZDCPos}
\end{figure}

The cross sections were measured for reference triggers based on two scintillator
arrays (VZERO)\cite{ALICE} in both \pp\ and \PbPb\ collisions, 
and two neutron zero degree calorimeters (ZDC)\cite{ALICE} in \PbPb\ collisions.

Fig.~\ref{fig-V0ZDCPos} shows a sketch of the VZERO and ZDC detectors.
As shown in the top panel, the VZERO consists of two scintillator arrays surrounding the beam pipe,
asymmetrically placed at each side of the ALICE interaction point (IP).
One array (VZERO-A) is located at 329~cm from the ALICE IP in one direction,
while the other (VZERO-C) is located at 86-88~cm (depending on the segment) in the opposite direction.
The corresponding pseudo-rapidity coverages are $2.8 <\eta < 5.1$ for VZERO-A and $-1.7 > \eta > -3.7$ for VZERO-C.
In each array the scintillator tiles are arranged in 4 (radial) $\times$ 8 (azimuthal) segments with
individual photomultiplier-tube readout.

With the VZERO, two different trigger types were prepared: VBAND (VZERO Beam {\it AND}) logic 
used in \pp\ mode; VLN (VZERO Low-threshold {\it AND}) logic used in \PbPb\ mode.

For VBAND, 32 signals from each array are discriminated and combined into a logical {\it OR}.
The two resulting signals are combined with an {\it AND} logic to have less sensitivity to background.
The coincidence window is 25~ns which corresponds to 10 LHC RF-buckets.

The VLN logic was configured to trigger on the $\simeq$50$\%$ most central hadronic Pb-Pb collisions.
At the front-end electronics, the integrated charge over 25~ns was measured for each of the photo-multiplier
tubes with proper pedestal value subtraction, and the sum of the integrated charges $Q_\mathrm{A}$ and $Q_\mathrm{C}$ were calculated,
individually on the VZERO-A and VZERO-C, respectively.
The innermost ring of VZERO-A was excluded in the processing for a technical reason.
The trigger condition is fulfilled if $Q_\mathrm{A}>T_\mathrm{A} \cap Q_\mathrm{C}>T_\mathrm{C}$ where $T_\mathrm{A}$ and $T_\mathrm{C}$ are threshold values
determined by a Glauber model fit to the VZERO total charge distribution in offline analysis\cite{CENTRALITY}.
The thresholds are different on the A and C side (roughly $T_\mathrm{C} \simeq 1.6\:T_\mathrm{A}$) due to differences
in detector geometries, construction, and secondary particle contamination.
At the selected set of thresholds, the trigger efficiency is above 95\% for collisions at 50\% centrality.
The expected cross section for the VLN logic is 52-53\% of the total Pb+Pb hadronic interaction cross section ($\sim$7.65~barn).

The ZDC detector consists of two sets of calorimeters, one positioned at $+114$~m and the other at $-114$~m
from the interaction point along the beam pipe direction and behind the intersection point of two beam lines,
as shown in Fig.~\ref{fig-V0ZDCPos}.
Each set of calorimeters consists of one proton ZDC and one neutron ZDC.
The neutron ZDCs are  located between two beam pipes.
In the present analysis, the proton ZDCs were not used and the reference trigger was configured as the logical {\it OR} of
two neutron ZDCs.
This trigger logic is used for EMD studies and thus called ZED.
The ZED trigger is sensitive to both nuclear interactions and EMD (both single and mutual)
where the latter dominates the rate.
The {\it AND} logic was not considered because  pile-up of single EMD will be significantly large and corrections
become non-trivial, although background effects are smaller.

\section{van der Meer Scan}

\begin{table}[t]
   \centering
   \caption{Overview of performed ALICE vdM scans. The scan ID is arbitrarily assigned and relevant only for this paper.
            Beam energies are given in TeV.
            $n_b$ is the number of colliding bunches at ALICE IP.
            The amplitude function $\beta^\ast$ is given in meters.
            Scan scheme details are given in the main text.
            The last column shows the status of analysis:
            F=Finalized, FG=Finalizing, NP=Not Planned, P=In Progress, S=Started.}
   \vspace{2mm}
   \begin{tabular}{lllll}
       \toprule
       \textbf{ID,}       & \textbf{beam,}& \textbf{fill,}                 &   \textbf{scheme}  & \textbf{stat.} \\ 
       \textbf{time}    & \textbf{$\sqrt{s}$}   & \textbf{$n_b$, $\beta^\ast$}&                 &                \\ 
       \midrule
           I,             & \pp           & 1090,                          & Xu-Xu              & F              \\
           May 2010       & 7             & 1, 2                           &                    &                \\
       \midrule
           II,            & \pp           & 1422,                          & Xu-Yu-             & F              \\
           Oct.2010       & 7             & 1, 3.5                         & Xd-Yd              &                \\
       \midrule
           III,           & \PbPb         & 1533,                          & Xu-Yu-             & F              \\
           Nov.2010       & 2.76          & 114, 3.5                       & Xu-Yu              &                \\
       \midrule
           IV,            & \pp           & 1634,                          & only test          & NP             \\
           Mar.2011       & 7             & 26, 10                         &                    &                \\
       \midrule
           V,             & \pp           & 1653,                          & Xu-Yu              & F              \\
           Mar.2011       & 2.76          & 48, 10                         &                    &                \\
       \midrule
           VI,            & \pp           & 1783,                          & Xu-Yu-             & FG             \\
           May 2011       & 7             & 16, 10                         & Xu-Yu              &                \\
       \midrule
           VII,           & \pp           & 2234,                          & Xu-Yu-Xd-          & S              \\
           Oct.2011       & 7             & 16, 10                         & Yd-Xuo-Yuo         &                \\
       \midrule
           VIII,          & \PbPb         & 2335,                          & Xu-Yu-Xu-          & P              \\
           Dec.2011       & 2.76          & 324, 1                         & Xd-Yd              &                \\
       \bottomrule
   \end{tabular}
   \label{table-scans}
\end{table}

\begin{figure*}[t]
    \centering
    \includegraphics*[width=150mm]{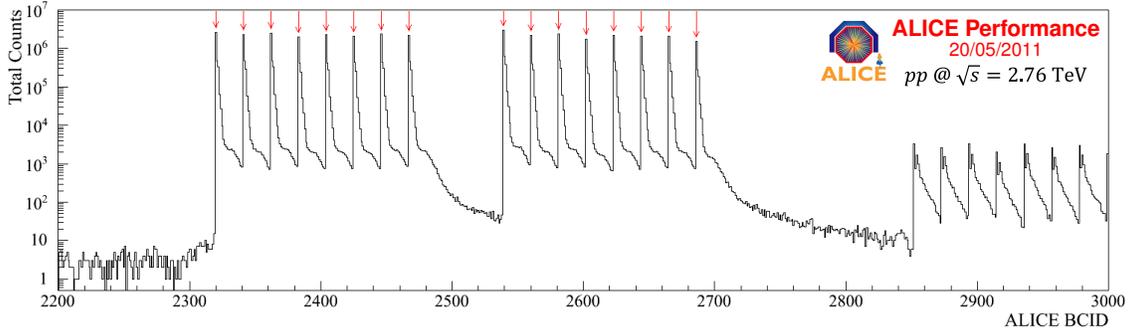}
    \caption{
    Interaction pattern during a vdM scan (Scan-V), as seen in ALICE. 
    The histogram shows the total integrated VBAND trigger counts and the arrows
    show the scheduled time for collisions expected for the LHC filling scheme.
    }
    \label{fig-BCIDdist}
\end{figure*}

In the vdM scan\cite{VDM}, the luminosity $\mathcal{L}$ and trigger rate are varied by changing the distance
between the two beams horizontally ($x$ direction) and vertically ($y$ direction), where the $x$-$y$ plane is 
perpendicular to the beam axis.
The $x$ and $y$ scans are performed individually.
The functional shapes of the trigger rate with respect to the displacements are obtained,
with several corrections which will be discussed in later sections.
The functional shapes of the rate with respect to the beam separation and their integrated area
($S_x$ and $S_y$) directly reflect the convolution of the transverse profiles of the two colliding
beams\cite{VDMOYAMA,VDMOYAMAQM,VDMGAGLIARDI}.
In \pp\ analysis, $S_x$ and $S_y$ are obtained as numerically calculated integrals of the separation versus rate graph.
Combining this information with the colliding bunch intensities $N_{1}$ and $N_{2}$
measured by beam instrumentation, the maximum luminosity at zero separation is calculated by
\begin{equation}\label{eq:cross}
  \mathcal{L} = \frac{f N_1 N_2}{ h_x h_y },
\end{equation}
where $f$ is the accelerator revolution frequency, and $h_{x,y}$ are the effective transverse widths of the colliding
beams in $x$ and $y$ direction, respectively.
The effective transverse widths are obtained as a ratio $h_{x,y} = S_{x,y} / R_{x,y}$,
where $R_{x,y}$ are the maximum trigger rate in each $x$ and $y$ scan.

During the vdM scan, the $N_{1,2}$ are monitored by the LHC beam instrumentation based on inductive current pickup devices.
The calibration of the instruments, correction on $N_{1,2}$, and evaluation of systematic uncertainties 
were performed by the beam current normalization working group (BCNWG) organized by the LHC and
experiments\cite{BCNWG-CERNREF,BCNWG-DCCT,BCNWG-GHOST,BCNWG-FBCT,ALICI}.

Table~\ref{table-scans} shows the summary of vdM  scans performed for ALICE in 2010 and 2011.
There were in total 8 scans performed with various beam setups including \PbPb.
In this paper, final results are reported for Scan-I, II, III, V and partial results for Scan-VIII.
Additionally, other scans such as Scan-VI and Scan-VII give important information on the systematics such as
reproducibility, hence, those are discussed in this paper too.

Fig.~\ref{fig-BCIDdist} shows a typical orbit structure of the LHC beams seen by the VBAND trigger during Scan-V.
The horizontal axis of the figure corresponds to the phase within the orbit in 25~ns steps,
called the bunch crossing identifier (BCID), with arbitrary offset.
The figure shows 22\% of a complete orbit which consists of 3564 BCID slots.
The snapshot corresponds to the time duration of the entire fill
and the vertical axis corresponds to the total number of trigger counts for a given BCID.
In Scan-V, there were 48 colliding bunch pairs per orbit at the ALICE IP.
Those are seen as large spikes at about $10^6$ counts indicated by arrows.
Smaller peaks at below $10^4$ counts are mainly due to beam-gas interactions corresponding
to unpaired bunches passing through the ALICE IP.
Beam-gas background contamination for bunch pair collision amounts to up to 0.2\%.
There is a long tail after each collision.
This is due to after-pulses of the photomultiplier tubes or after-glow
due to slow particles and the activation of the surrounding materials.

In analysis, such snapshots of trigger counters were produced at each beam separation. Thus,
the measurement of the rates, beam geometries, luminosities, and cross sections were
performed pair-by-pair for all colliding bunch pairs.

\section{Corrections}

The obtained rate data during the vdM scan had to be corrected for various effects.
The corrections may depend on beam separation and luminosity, or other effects.
There are additional corrections to be globally applied to the measured cross sections.
In this section, details of those corrections are given.

\subsection{Pile-up corrections}

\begin{figure}[tb]
    \centering
    \includegraphics*[width=55mm]{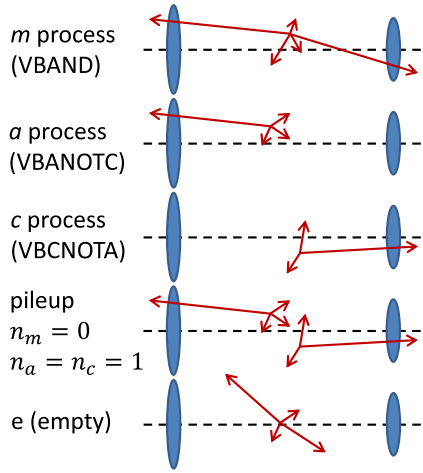}
    \caption{
    Top three drawings show the processes (\mbox{$m$-,} \mbox{$a$-,} and \mbox{$c$-processes})
    detected by different trigger conditions.
    Fourth drawing shows an example configuration for pile-up events where there is no $m$-process
    but one for each $a$- and $c$-processes thus fulfills $m$-process trigger condition.
    The last one is $e$ which means empty. Empty does not necessarily mean that there is no interaction.
    }
    \label{fig-pileup}
\end{figure}

A bunch crossing in which more than one interaction occurs is still counted as one trigger fired (pile-up effect).
Assuming Poisson statistics, the trigger rate is reduced by factor
\begin{equation}\label{eq:simplepileup}
 ( 1-e^{-\mu} )/\mu,
\end{equation}
where $\mu$ is the average number of interactions per bunch crossing.
The correction factor using Eq.\ref{eq:simplepileup} is up to 40\% and dominates for small displacement
and at high luminosity.
In the previous analysis of Scan-I, only this type of simple pile-up was considered\cite{VDMOYAMA,VDMOYAMAQM,VDMGAGLIARDI}.
The VBAND condition requires at least one charged particle hit on both VZERO arrays, and thus requires a particle multiplicity of 2 or more.
The above correction method considers pile-up of events with two or more tracks in VZERO.
However, there is another type of pile-up.
If two single track events occur simultaneously, and one hits the A side and the other hits the C side,
then such an event is counted as a a single event although each separately does not fulfill the trigger condition.

The probability of this pile-up is considered to be small and its effect is
negligible compared to the total systematic uncertainty in Scan-I. 

However, as the measurements became more precise, more exact pile-up corrections were needed for Scan-II and later.
Fig.~\ref{fig-pileup} illustrates the configurations of processes that can occur in one bunch crossing
and that are detected by VZERO.
The top three configurations are all single events without pile-up.
For convenience, these are called ($m$, $a$, and $c$-processes).
Without pile-up, only the $m$-process can fulfill the VBAND condition.
Process $a$($c$) can fulfill the VBANOTC(VBCNOTA) condition which means that only the A(C) side
of the VZERO detector detects particles.
The fourth configuration shows an example pile-up events which was not considered in the simple pile-up correction
method using Eq.\ref{eq:simplepileup}.
There can be invisible events not detected by VZERO (last case).
However, such events do not affect the pile-up correction.
Assuming again Poissonian probability for the occurrence of each process,
the corrected average number of processes in one bunch crossing $\mu_m$, $\mu_a$, $\mu_c$
can be obtained from the raw trigger rates $R_m^\mathrm{t}$, $R_a^\mathrm{t}$, and $R_c^\mathrm{t}$ of the bunch-pair being analyzed by:
\begin{equation}\label{eq:precisepileup}
  e^{-\mu_m}=\frac{ (1-R_m^\mathrm{t}/f - R^\mathrm{t}_a/f ) ( 1-R_m^\mathrm{t}/f - R_c^\mathrm{t}/f )}
                  { 1-R_m^\mathrm{t}/f - R_a^\mathrm{t}/f - R_c^\mathrm{t}/f }
\end{equation}
and
\begin{equation}
  e^{-\mu_{a,c}}=\frac{ 1-R_m^\mathrm{t}/f - R_a^\mathrm{t}/f - R_c^\mathrm{t}/f }
                    { 1-R_m^\mathrm{t}/f - R_{c,a}^\mathrm{t}/f},
\end{equation}
where $f$ is the constant accelerator revolution frequency.
In the present case $R_m^\mathrm{t}$ is the rate of VBAND, and $R_a^\mathrm{t}$ and $R_c^\mathrm{t}$  are rates of the VBANOTC and VBCNOTA conditions,
respectively.
The fully corrected process rates $R_m$, $R_a$, and $R_c$ are calculated using:
\begin{equation}
  R_{m,a,c} = -f \ln e^{-\mu_{m,a,c}}
\end{equation}
It should be noted that in the case of the ALICE VZERO, the process $a$ and $c$ cross sections and therefore their
rates are not equal because the VZERO acceptance is asymmetric.

The ratios of the cross sections ($\sigma_{a,c,m}$) of these processes
$\mu_a/\mu_m=\sigma_a/\sigma_m$ and $\mu_c/\mu_m=\sigma_c/\sigma_m$ are constant within statistical uncertainties.
For the Scan-II case, $\mu_a/\mu_m\sim0.08$ and $\mu_c/\mu_m\sim0.07$ over a wide range of separations.
The estimated cross sections for $\sigma_a$ and $\sigma_c$ obtained from these ratios together with finally
measured VBAND cross section are 4.3~mb and 3.6~mb, respectively, for 7~TeV \pp\ collisions.

Corrected rates by the method using Eq:~\ref{eq:precisepileup} is 0.35\% smaller than the rates corrected 
using Eq:~\ref{eq:simplepileup}.
The overall effect on the cross section is 0.16\% for the Scan-II case.

\subsection{Luminosity Decay Correction}

Typically, one set of vdM scans takes about 30 minutes.
During this time, the beam conditions change due to decreasing bunch intensities and increasing beam emittances. 
This is seen as a luminosity decay, and the shape of the trigger rate versus beam separation is deformed by this effect.
To correct for this decay, the rate data at each separation was normalized to the rate at a given reference time.
Typically, the middle of the horizontal and vertical scan time was used as a reference time.
The decay slope obtained by a straight line fit for the trigger rate just before and just after the scans
was used to normalize the data rate.
The method is the same since the first analysis was performed and reported\cite{VDMOYAMA}.
The correction factor depends on the beam situation and the typical maximum correction amounts to 2 to 3\%.

\subsection{Background and Satellite Correction}

In a typical LHC optics setup, the crossing angle of the two beams is in the vertical plane at the ALICE IP.
As the beam separation increases in the vertical direction during the vdM scan, 
the two main bunches being measured move away from each other.
However, the satellite charges in the next RF bucket ($\pm 2.5$~ns away) or the following RF
bucket ($\pm 5.0$~ns away) of the main bunch start to collide with the main bunch of the other beam.
These are satellite-main collisions and they enhance the trigger rate at large separation.
The satellite-main collisions happen 1.25~ns or 2.50~ns later or earlier in time compared to main-main collisions.

In addition, beam-gas collisions in the beam pipe can create fake triggers.
Most of the beam-gas collisions take place far away from the ALICE detector.

\begin{figure}[tb]
    \centering
    \includegraphics*[width=75mm]{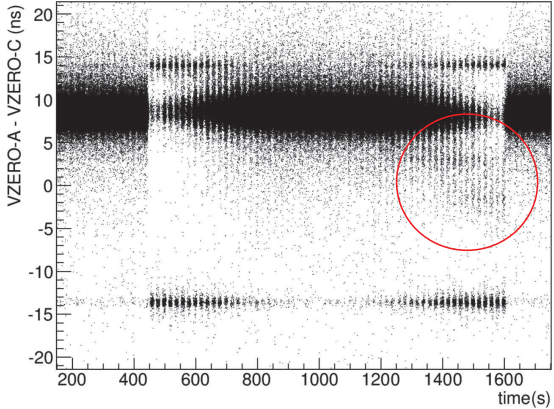}
    \caption{
      VZERO-A and VZERO-C timing difference with respect to the time when vertical scan has been performed.
      This is an example of Scan-V.
    }
    \label{fig-satback}
\end{figure}

The background effects due to both satellite-main collisions and beam-gas background events affecting the measured trigger rates
can be removed by checking the particle arrival time at VZERO-A and VZERO-C.
Fig.~\ref{fig-satback} shows the difference of the pulse arrival time between VZERO-A and VZERO-C
with respect to the time when the vertical vdM scan was performed.
The time axis has arbitrary offset and $\sim$450~s and $\sim$1600~s correspond to
the time when the beams had the maximum separation in opposite directions to each other.

The large magnitude band that is always present at 8~ns corresponds to main-main collisions.
Narrow bands seen at $\pm 14$~ns correspond to beam-gas events while
the less frequent entries indicated by a circle correspond to satellite-main collisions.
The relative rate of beam-gas events becomes smaller for head-on collisions
(at $\sim$1000~s) due to less trigger live-time for data recording.

Using this analysis result, histograms were made for each separation value, and the relative
abundance of the main-main collisions with respect to the total trigger rates were calculated for each
separation using a Gaussian fit approach.
At the maximum separation, for Scan-V, the fraction of main-main collision was 30\% of the total events.
The situation changes with each scan.
All the other scans have larger fraction of main-main collisions.

\subsection{Ghost and Satellite Charge Correction}

There are charges distributed over the accelerator orbit, and also satellite charges
associated to the main bunches.
The beam intensity data provided by BCNWG are overestimated because those are not corrected
for ghost and satellite charges thus leading to an under-estimation of the cross sections.
The correction factors were separately provided by BCNWG as well.
These are typically below 1\% except for Scan-V where the correction factor for the charge product
of two beams becomes 2.5\%.

A correction is needed also for the satellite charge (i.e. charge populating a non-nominal RF bucket within
the colliding bunch slot).
The satellite charge can be estimated by ALICE\cite{BCNWG-DCCT} using the VZERO pulse arrival time
as in Fig.~\ref{fig-satback}.
Such a correction was found to be non-negligible for Scan-II ($\sim$0.8\%).

\subsection{Length Scale Calibration}

\begin{table}[tb]
   \centering
   \caption{Length scale calibration results. Correction factors for $x$ and $y$ direction of beam displacement are listed.}
   \vspace{2mm}
   \begin{tabular}{llll}
       \toprule
       \textbf{time}      & \textbf{Fill} & \textbf{$x$ corr.}  & \textbf{$y$ corr.} \\
       \midrule
       May  2010          & 1090          & 0.987             & 0.991             \\
       Oct. 2010          & 1455          & 1.003 $\pm$ 0.010 & 0.996 $\pm$ 0.005 \\
       Mar. 2011          & 1658          & 0.993 $\pm$ 0.002 & 0.995 $\pm$ 0.003 \\
       May  2011          & 1783          & 0.992 $\pm$ 0.003 & 0.997 $\pm$ 0.003 \\
       Dec. 2011          & 2335          & 0.993 $\pm$ 0.005 & 1.016 $\pm$ 0.004 \\ 
       \bottomrule
   \end{tabular}
   \label{table-lsc}
\end{table}

The scale of the spatial separation of the two beams during the vdM scan has been calibrated by taking data when
both beams were moved to the same direction instead of the opposite directions, and measure the movement of the
collision vertex (length scale calibration: LSC).
The movement of the event vertex distribution was measured by the Silicon-Pixel Detector.
Table~\ref{table-lsc} summarizes the LSC operations carried out in ALICE and the results.
The LSC is not necessarily performed in the same fill as the vdM scan.
The result of LSC performed without changing the accelerator optics configuration
can be still used for the corrections in corresponding vdM scan.

\subsection{Summary of Corrections}

\begin{figure*}[t]
    \centering
    \includegraphics*[width=75mm]{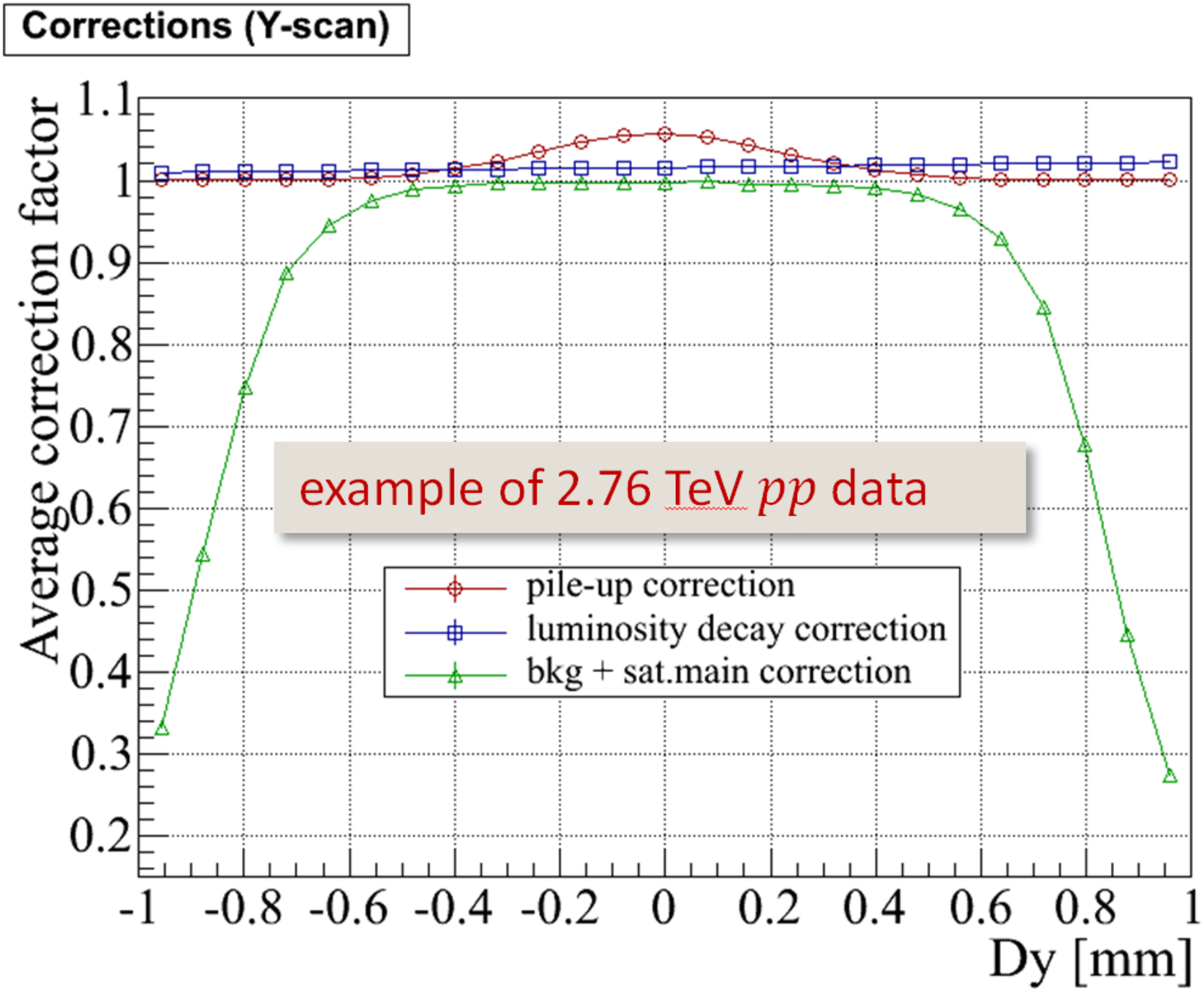}
    $\;\;\;\;\;\;$
    \includegraphics*[width=54mm]{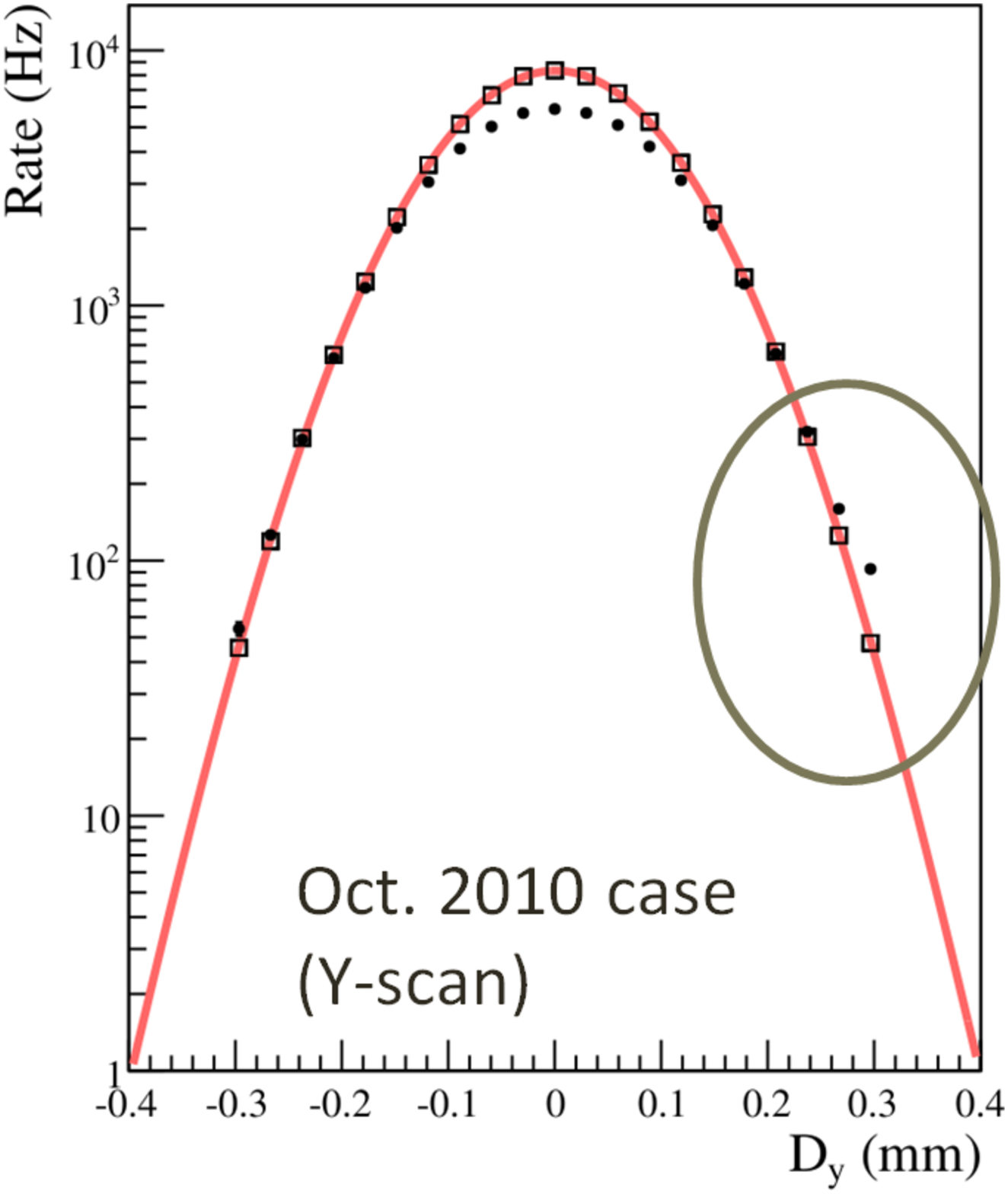}
    \caption{
      Left: Beam separation dependent correction factors applied for Scan-V analysis for Scan-V vertical scan case case.
      Right: Comparison of rates before and after the corrections for Scan-II vertical scan case.
      Solid circles are before any corrections, and open boxes are after applying all the corrections.
      The Gaussian fit (red solid line) was performed only to guide eyes,
      and a circle indicates that there is large effect of satellite and background at large displacement.
    }
    \label{fig-corrections}
\end{figure*}

Fig.~\ref{fig-corrections} shows the correction factors applied for Scan-V analysis (left) and typical data
before and after correction for Scan-II (right), only for the vertical separation case.
Only corrections depending on the separation (pile-up correction, luminosity decay correction,
and background and satellite corrections) are shown in the right plot.
The correction factors are to be used to divide the obtained raw trigger rates to get the corrected rates at a given
beam separation.
The largest correction at small separation is the pile-up correction while the background and satellite correction becomes
dominant at large separation as seen on both plots.

Table~\ref{table-corrections} summarizes the global correction factors
(ghost charge, satellite charge, and LSC corrections) applied for Scan-II, V, and VI analysis.

\begin{table}[tb]
   \centering
   \caption{Summary of global correction factors. The given values should be multiplied
            with cross section value to arrive at the corrected value.}
   \vspace{2mm}
   \begin{tabular}{llll}
       \toprule
       \textbf{ID}   & \textbf{ghost}    & \textbf{satellite}  & \textbf{LSC} \\
                     & \textbf{charge}   & \textbf{charge}     & \\
       \midrule
       II            & 0.9925$^{-1}$     & 0.992$^{-1}$        & 0.9986 \\
       V             & 0.9748$^{-1}$     & negligible          & 0.9886 \\
       VI            & 0.9966$^{-1}$     & not yet available   & 0.9894 \\
       \bottomrule
   \end{tabular}
   \label{table-corrections}
\end{table}

\section{Systematic Uncertainties}

\begin{figure*}[tb]
    \centering
    \includegraphics*[width=80mm]{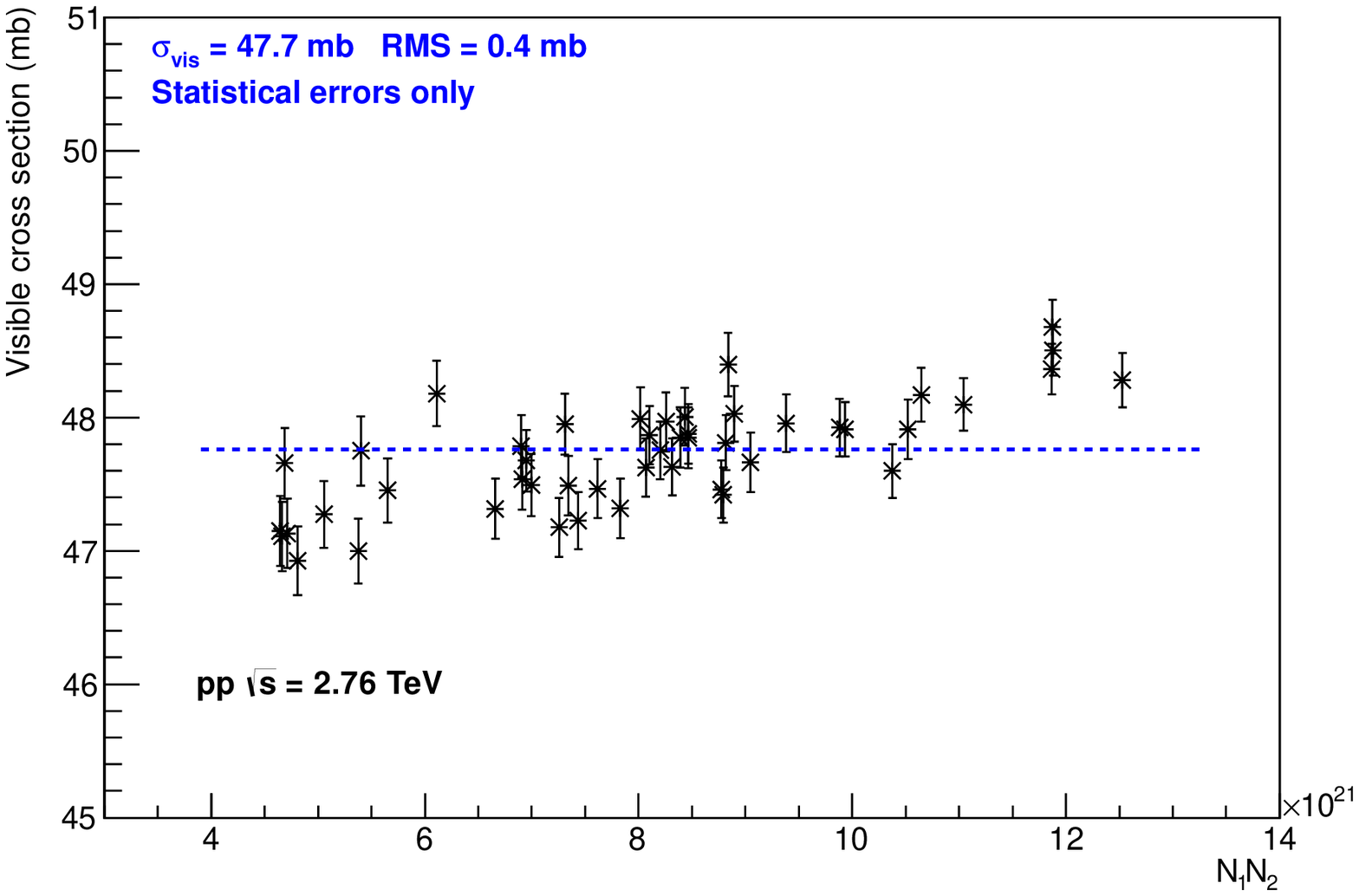}
    \includegraphics*[width=80mm]{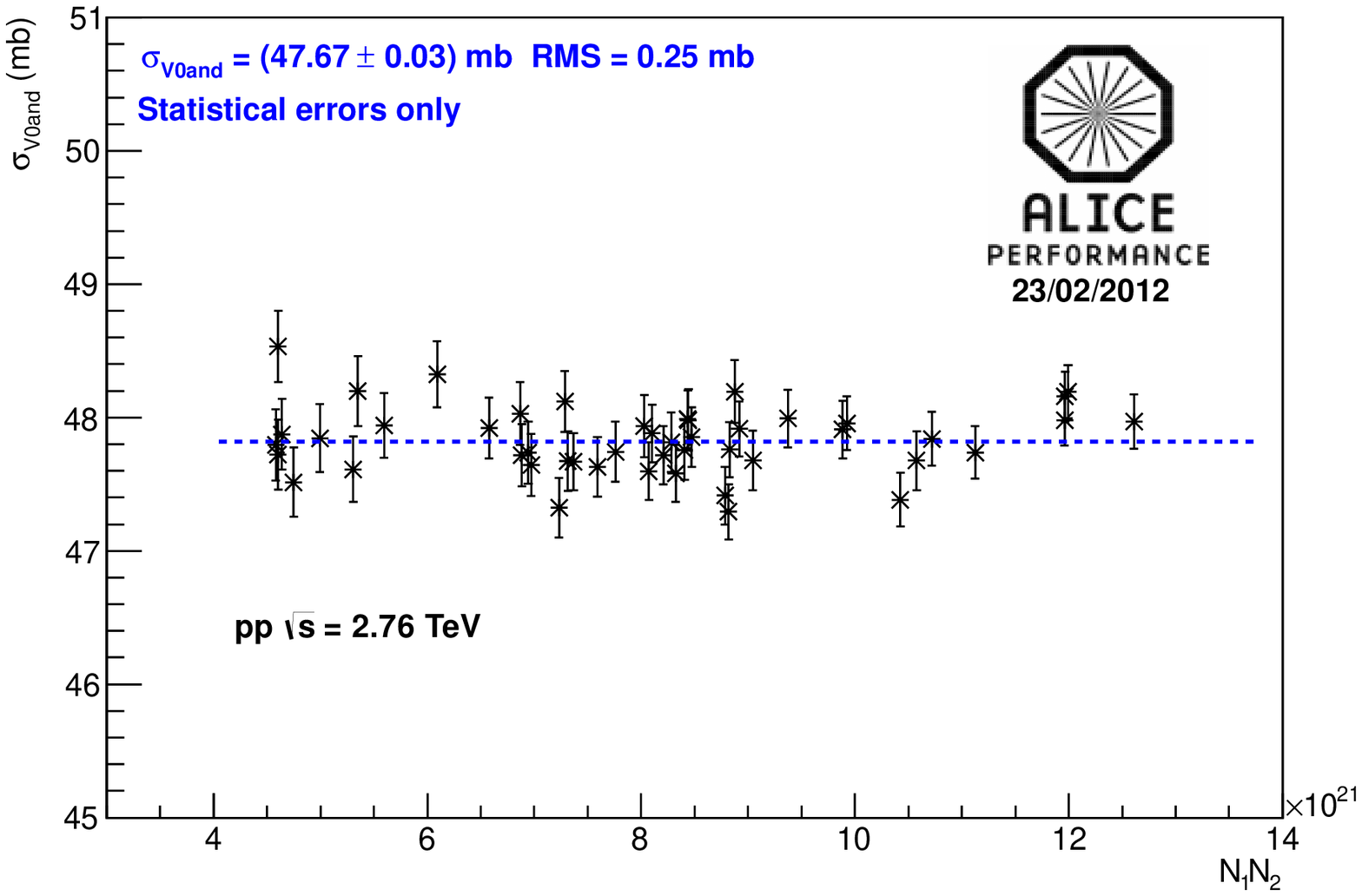}
    \caption{
      VZERO cross section as functions of bunch intensity product $N_1N_2$, before the offset correction (left)
      and after the offset correction (right).
    }
    \label{fig-offset}
\end{figure*}

In this section, the systematic uncertainties and reproducibility of the results
within the same accelerator fill or over different fills, and other possible effects are discussed.

\subsection{Bunch-by-Bunch Comparison}

Until Scan-IV, the data acquisition  system of ALICE was not able to individually measure trigger rates for more than one colliding
bunch pairs at the same time.
The multi-bunch analysis and comparisons became available from Scan-V.
In Scan-V analysis, there were 48 colliding bunch pairs and those were individually analyzed and provided a systematic
comparison of the cross sections in wide ranges of bunch intensities and geometrical beam sizes.

Fig.~\ref{fig-offset} shows the comparison of cross sections with respect to the 
bunch intensity product ($N_1 N_2$) measured in Scan-V.
In the left panel of the figure, an RMS spread of 0.8\% can be seen,
with a clear correlation to the bunch intensity product.
The deviation is small compared to the variation of the beam intensity and beam spot sizes.
However, a clear dependency on the bunch intensity product is seen.
This dependency is explained as due to a residual offset in bunch intensity measurements.
The offset value was obtained by fitting with the value as a free parameter.
As shown in the right panel of Fig.~\ref{fig-offset}, the correlation to the bunch intensity product disappears
after such an offset correction.
The residual RMS spread is reduced to 0.5\% which is at the level of statistical fluctuation.

For multi-pair analyses, the cross section values are averaged over all colliding bunch pairs.
This method also has the advantage that the systematic uncertainty for bunch intensity measurements, 
which are uncorrelated between bunches, vanishes by averaging.
After averaging for Scan-V results, the cross section values essentially do not differ before
and after offset correction.
Thus, this correction was not necessary in multi-pair analyses.

\subsection{Reproducibility Check}

\begin{figure*}[tb]
    \centering
    \includegraphics*[width=150mm]{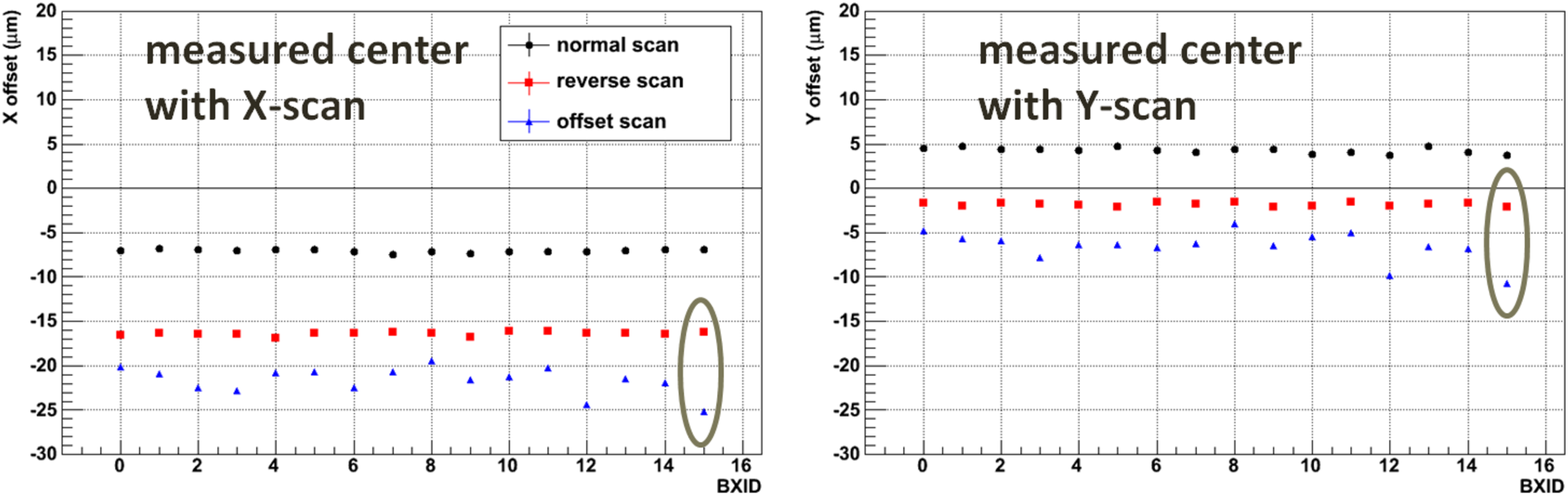}
    \caption{
      Measured center position of the offset scans in $x$ direction (left) and $y$  direction (right).
      The numbers in the horizontal axis is arbitrary given to identify the colliding bunch pairs.
      Black solid circles and red solid boxes are data taken without offset, corresponding to the first (normal) scan
      and the second (reverse) scan, respectively.
      Blue solid triangles correspond to offset scan.
    }
    \label{fig-center}
\end{figure*}

In Scan-II, as shown in Table~\ref{table-scans}, two scan sets were obtained.
The first one is a Xu-Yu pattern where the separation of the beams was ``increased'' while
in the second set, the pattern Xd-Yd was performed where the beam separation was ``decreased''
by moving in reverse direction with respect to to the first scan pattern.

Those two scan sets were analyzed independently, and it turned out that the resulting cross
section values differ by 0.8\%.
This is a factor of 2 larger than the possible differences due to the statistical uncertainty (0.4\%)
and is thus considered as a systematic difference due to unreproducibility.
This is under investigation with further scan data and possible reason is the presence of magnet hysteresis.
Since the observed difference is peak-to-peak, half of the difference is assigned as 
systematic uncertainty which is summarized later.

The average of these two cross section values was used as the central value of the Scan-II result.

The similar check has been performed for Scan-VI
when there were two sets of scans performed but both were ``increased'' scan pattern.
In this case, a 0.16\% discrepancy between the two scan sets was observed.
This is significantly smaller than the Scan-II case.

\subsection{$x$-$y$ Coupling}

The vdM scan and its analysis is based on the assumption that the bunch shape at the interaction point
is represented by $\rho(x,y)=Q p(x)q(y)$ where $\rho$ is the charge density distribution in the $x$-$y$ plane
perpendicular to the beam axis, $Q$ is the total bunch charge, and $p$ and $q$ are the normalized
distribution in $x$ and $y$, respectively\cite{VDMOYAMA}.
If this factorized beam shape assumption is far from realistic and there is a coupling
between $x$ and $y$ directions, the analysis results in a wrong cross section value.
This can happen if the beam shape is oval and there is a residual angle between the oval axis
and beam separation direction due to the way the accelerator was built.

This $x$-$y$ coupling effect can be measured if there are two scans performed in the same plane,
for example in $x$-direction, but keeping the offset separation in the other plane (``offset scan'') during one of the scans.
The first offset scan was attempted in Scan-VII
when there were three sets of scans taken as shown in Table~\ref{table-scans}.
After two sets of scans with nominal Xu-Yu scan pattern and Xd-Yd (reverse) scan pattern,
the third scan set with offset gives a Xuo-Yuo pattern shown in the table.
While the Xuo scan was taken, the two beams were separated also in the $y$ direction by $\sim$1.5$\sigma$.
Since there were other activities at other interaction points between the first and second scan sets,
there was more than 3 hours between the first and second scan while the third scan was performed
right after the second scan.
Therefore due to orbit drifts, there is a possible change in the beam position between the first and the second scan.
However, there should be much less drift between the the second and the third scan if there is no $x$-$y$ coupling.

Fig.~\ref{fig-center} shows the results of those scans.
The beam center positions were obtained by fitting a Gaussian function to the scan data, for each crossing pair.
In $x$-scans (left panel), a large drift of $\sim$10~$\mu$m is seen from the first Xu  scan (black)
to the second Xd scan (red) which can be interpreted as a natural orbit drift.
In addition, there is a further change in the second and third Xuo scans (blue) by $\sim$5~$\mu$m and
the fluctuation from bunch-pair to bunch-pair seems larger although it was performed immediately
after the previous scan.
This may indicate the presence of $x$-$y$ coupling and tilted beams with the situation different from bunch to bunch.
A quite similar structure is seen also in the Yuo scan (right panel), again indicating tilted beams.
It should be noted, that for the same bunch pairs indicated by circles, the largest variation of $\sim$10~$\mu$m
is seen both in Xuo and Yuo scans.
In the present analysis, it was considered that there will be  $\sim$10~$\mu$m effect of $x$-$y$ coupling
to stay on the safe side, although the observed maximum change may still include beam drift.
The impact on the final cross section value was estimated to be 0.3\% for each direction using a model calculation.
Since the effects are totally correlated in $x$ and $y$ directions, the total effect is 0.6\%
which was introduced as a systematic uncertainty.

\subsection{Summary of Systematic Uncertainties in \pp}

Table~\ref{table-syst} summarizes the systematic uncertainties considered for Scan-II and Scan-V results in \pp\ collisions.

\begin{table}[tb]
   \centering
   \caption{Summary of relative systematic uncertainties for analyzed \pp\ scans.
     ``NC'' in the table indicates that this effect was not considered at the moment the analysis is performed.
   }
   \vspace{2mm}
   \begin{tabular}{lcc}
       \toprule
       \textbf{item}                    & \textbf{Scan-II} & \textbf{Scan-V}   \\
       \midrule
       bunch intensity $\delta(N_1N_2)$ & 3.2\%            & 0.57\%            \\
       \midrule
       length scale calibration         & 1\%$\oplus$1\%   & 1\%$\oplus$1\%    \\
       luminosity decay                 & neglig.          & 0.5\%             \\
       hysteresis, reproducibility      & neglig.          & 0.4\%             \\
       beam centering                   & neglig.          & neglig.           \\
       after-pulse / after-glow         & neglig.          & 0.2\%             \\
       background \& satellite rate     & neglig.          & 0.3\%             \\
       pile-up correction               & neglig.          & neglig.           \\
       $x$-$y$ coupling                 & NC               & 0.6\%             \\
       dynamic $\beta^\ast$             & NC               & 0.4\%             \\
       \midrule
       total in experiment              & 1.41\%           & 1.75\%            \\
       total with bunch intensity       & 3.50\%           & 1.84\%            \\ 
       \bottomrule
   \end{tabular}
   \label{table-syst}
\end{table}

Beam intensity uncertainties are provided by BCNWG.
The estimated beam intensity uncertainty for Scan-II is 3.2\%.
Scan-V has a smaller beam intensity uncertainty of 0.57\% comprising the uncertainties on ghost charge
estimation (0.4\%)\cite{BCNWG-GHOST}, total beam current measurement by DC current transformer (DCCT)
at 0.34\%\cite{BCNWG-DCCT},
relative bunch population measurement by fast bunch current transformer (fBCT) at 0.08\%\cite{BCNWG-FBCT},
and satellite charge effect (0.2\%)\cite{SATELLITE}.

The LSC results showed that the length scale is satisfactorily stable,
however, conservative values of 1\% are assigned for each scan direction.
Future studies with many vdM scans may give more precise values.

The luminosity decay corrections result in an up to 1\% effect on the cross section value.
Half of this effect was assigned as a systematic uncertainty.

For hysteresis and reproducibility, half of the observed maximum difference already presented in the
{\it Reproducibility Check} section is assigned.
For Scan-II, it is negligibly small compared to other dominant uncertainties.
For Scan-V it is of considerable significance.

If beams are misaligned in one plane while scanning the other plane, the observed luminosity and
trigger rate will be reduced and the cross section obtained will be underestimated.
The misalignment is seen as a shift of the beam shape for each direction scan,
and a correction can be performed.
In the cases of Scan-II and Scan-V this turned out to be either negligibly small compared to
other effects, or zero.

The after-pulse and the after-glow may give a higher trigger rate.
Since the after-pulse can occur randomly up to 1~$\mu$s from the main pulse, an after-pulse alone
does not fulfill coincidence condition of VBAND.
However, pile-up of the after-pulse, together with a single track physics event on the other side of VZERO,
will contribute as fake trigger rate in VBAND.
This fake trigger rate was estimated using the observed after-pulse or after-glow magnitude,
and using exclusive events ($a$- and $c$-processes) cross section described already, to be a 0.2\% effect.
Instead of correcting data, the 0.2\% was assigned as systematic uncertainty.

The background and satellite rate correction was discussed in the previous section.
Although the correction factor is significant at large separation for the Scan-V case,
the overall effect on the cross section is 0.6\%, and half of this value was assigned as systematic uncertainty.

After checking the results with different beam intensities and profiles, the pile-up
correction was shown to be well within limits.
No source of systematic uncertainty was found.

For the $x$-$y$ coupling, as discussed in previous subsection, the worst possible case was chosen
and assigned as systematic uncertainty (0.6\%).

The transverse sizes of the beam profile are not stable if the separation between the two colliding beams
is changed during the vdM scan because of beam-beam interactions.
The beams unfocus each other and the effective $\beta^\ast$ function will become smaller at
head-on collisions compared to larger separation (dynamic $\beta^\ast$ effect).
This leads to a change of the beam width during the scan.
The amount of such an effect was quantitatively estimated by calculations using a beam optics simulation program
(MAD-X) together with additional elements of the beam-beam interactions\cite{DYNAMIC-BETA}.
The calculation shows that $\beta^\ast$ decreases by up to $\sim$1\% during the scan.
This causes about a 0.2\% of modification in beam width and 
the effect on the cross section is 0.4\% the full value of which was conservatively introduced as a systematic uncertainty.

\section{Results for \pp\ collision}


Table~\ref{table-centralvals} shows the results of four vdM scans in $pp$ collisions analyzed in ALICE.
For 7~TeV, three scans (Scan-I, finalized), (Scan-II, finalized), and (Scan-VI, preliminary) are compared.
For 2.76~TeV the Scan-V result is given.
For the analysis of 2010 data at 7 TeV, the best estimation among all scans at this moment is Scan-II, $\SV$=54.34~mb,
in very good agreement with the result of Scan-I.

The stability of the VBAND cross section throughout 2010 run was further cross-checked by observing the stability of
the relative rate of the VBAND trigger with respect to other trigger rates such as a VBOR trigger
(logical {\it OR} of VZERO-A and VZERO-C, instead of logical {\it AND})
and a forward muon trigger using the ALICE muon trigger system.
The fluctuations in the ratio were found to be negligible with respect to the other sources of uncertainty in the
VBAND cross section\cite{VDMOYAMA}.

The result from Scan-VI (performed in 2011) exhibits a 1.2\% smaller cross section.
This is under investigation and one of the possible reasons for this is the ageing of the VZERO detector.
If confirmed, such effect will have to be taken into account in the luminosity determination for 2011 data.

\begin{table}[tb]
   \centering
   \caption{Comparison of central values for cross section among \pp\ vdM scans.
            It should be noted that Scan-V has different collision energy than others.
            (*) Systematic uncertainty for Scan-VI is not yet available.
   }
   \vspace{2mm}
   \begin{tabular}{lcccc}
       \toprule
       \textbf{ID}             & \textbf{I} & \textbf{II}  & \textbf{V} & \textbf{VI} \\
       \midrule
       $\sqrts$ (TeV)        & 7               & 7     & 2.76  & 7 \\
       $\SV$(mb)             & 54.21           & 54.34 & 47.67 & 53.67 \\
       diff. to Scan-II      & $-0.2$\%       & NA & NA & $-1.2$\% \\
       stat. uncert.         & $<$1\%          & 0.2\%  & 0.1\%  & 0.1\% \\
       syst. uncert.         & 7\%             & 3.5\%  & 1.84\% & (*) \\
       \bottomrule
   \end{tabular}
   \label{table-centralvals}
\end{table}

\section{Analysis in \PbPb\ Collisions}

In this section the analyses of the Pb-Pb vdM scans are briefly described.
The basic procedure of vdM scans in \PbPb\ is not different from that in \pp\ except that
the neutron ZDC is used in \PbPb\ in addition to the VZERO detector.

\begin{figure}[tb]
    \centering
    \includegraphics*[width=82mm]{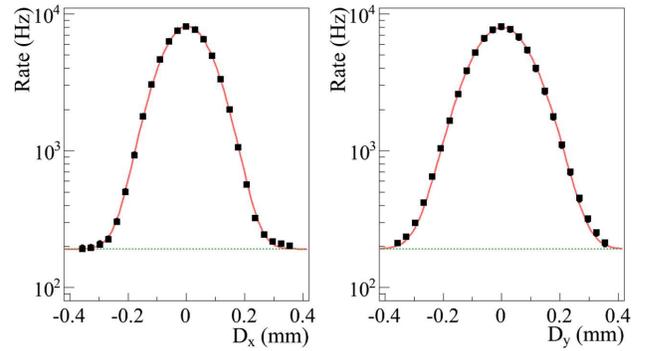}
    \caption{
    Corrected vdM scan shape, obtained in Scan-III with \PbPb\ beams, for ZED trigger.
    On the left is the horizontal scan, on the right the vertical.
    The solid line shows the fit with Gaussian plus baseline, and the dotted line shows only
    baseline component of the fit function.    
    }
    \label{fig-shapePbPb}
\end{figure}

In Scan-III, two sets of scans with pattern Xu-Yu were carried out, as shown in Table~\ref{table-scans}.
Analyses have been performed for both sets, and results were compared.
Fig.~\ref{fig-shapePbPb} shows example shapes of the ZED trigger rates with respect to beam separations
measured in the first set of Scan-III.
In contrast to the VZERO case in \pp, which had essentially no background at large separations,
the neutron ZDC has relatively high background and thus the baseline determination becomes important.
It was found that the separation is up to 0.38~mm for both $x$ and $y$ scans.
This is not enough to accurately determine the baseline.
The fit procedure with a Gaussian fit function plus a baseline is thus unstable due to the residual
tail.
In this analysis, the baseline was fixed at 191.6~Hz and 191.2~Hz for the first and second set of scans,
respectively.
These values were chosen as ``minimum'' observed rates during the scan.
The true background components might be different from the fixed value, and this was considered as a
source of systematic uncertainty, as will be discussed later.

There were 114 colliding bunch-pairs in Scan-III.
However the bunch-pair by bunch-pair analysis was not possible at this stage.
Only the inclusive rate over the orbit was available from the ALICE trigger system.
For this reason, the analysis was performed inclusively.

Since the obtained curve of the trigger rate with respect to the beam separation has Gaussian shape,
the result of the fit was directly used to determine the reference cross section by:
\begin{equation}\label{eq:inclusive-cs}
  \sigma = \frac{2\pi\sigma_x\sigma_y R^{\mathrm{top}}}{f\sum_{i=1}^nP_i}
\end{equation}
where $\sigma_{x,y}$ are the width obtained by Gaussian fits, $R^\mathrm{top}$ is the head-on trigger rate,
and $P_i$ is the product of intensities of two beams in $i$'th bunch pair.
These values are shown in Table~\ref{table-pbpb}, which summarizes the \PbPb\ scan parameters
and the results for the central value of the ZED cross section, individually for the first and second scan sets.

There is a discrepancy of about 0.9\% in the ZED cross sections between two scan sets.
This is larger than the statistical uncertainty of $\sim$0.2\%, and indicates a possible systematic effect
such as magnet hysteresis.
Therefore, the average of the two sets was used as central value.

\begin{table}[tb]
   \centering
   \caption{Scan-III result with obtained scan parameters.
   Only statistical uncertainties are given, except for sum of $N_1N_2$.
   Ghost charge measurement correction (for charge outside $\pm$15~ns window) is already applied.
   }
   \vspace{2mm}
   \begin{tabular}{lccc}
       \toprule
       \textbf{items}         &   & \textbf{first set} & \textbf{second set}  \\
       \midrule
       $\sum_{i=1}^nP_i$      &   & $6.814 \times 10^{21}$ & $6.079 \times 10^{21}$ \\
       $R^\mathrm{top}$ (Hz)  &$x$& 7946 $\pm$ 15      &  6012 $\pm$ 11  \\
                              &$y$& 7775 $\pm$ 14      &  6143 $\pm$ 12  \\
       $\sigma_{x,y}$ ($\mu$m)&$x$& 83.6 $\pm$ 0.1     &  104.2 $\pm$ 0.1 \\
                              &$y$& 99.1 $\pm$ 0.1     &  92.8 $\pm$ 0.1 \\
       center ($\mu$m)        &$x$& 2.8 $\pm$ 0.1      &  2.9 $\pm$ 0.2 \\
                              &$y$& 2.1 $\pm$ 0.2      &  -2.5 $\pm$ 0.2 \\
       baseline (Hz)          &$x$& 191.6 (fix)        &  191.2 (fix) \\
                              &$y$& 191.6 (fix)        &  191.2 (fix)\\
       $\SZED$ (barn)         &   & 369.7 $\pm$ 0.8       & 373.0 $\pm$ 0.9\\
       \midrule
       $\SZED$ avg. (barn)    &   & \multicolumn{2}{c}{371.4 $\pm$ 0.6} \\
       \bottomrule
   \end{tabular}
   \label{table-pbpb}
\end{table}

The VLN cross section $\SVLN$ was measured only in Scan-VIII where the bunch-by-bunch analysis was possible and
therefore all the 324 colliding pairs were individually used to measure $\SVLN$.
The individually measured cross section was averaged over all colliding bunch pairs by:
\begin{equation}\label{eq:bbb-cs}
  \sigma = \frac{1}{n} \sum_{i=1}^n \frac{2\pi\sigma_x^i\sigma_y^i R_{i}^\mathrm{top}}{f P_i}.
\end{equation}
Preliminary, $\SVLN$ was calculated to be 3.89 barn without any correction.
%
%
%
According to measurements made by BCNWG\cite{JEFF-GHOST}, 2.4$^{+2.4}_{-0.6}\:$\% and 2.1$^{+2.1}_{-0.5}\:$\%
ghost charges were observed for Beam-1 and Beam-2, respectively in the beam fill of Scan-VIII.
With ghost charge, the previously measured $\SVLN$ is underestimated by a factor of $(1-2.4\%)\cdot(1-2.1\%)=0.955$,
thus, the corrected $\SVLN$ together with LSC correction is 4.10 barn.
The statistical uncertainty is negligibly small.
The result is consistent within systematic uncertainty to the expected value of 4.0-4.1~barn,
found by combining the 7.65 barn Pb-Pb hadronic interaction cross section with 52-53\% trigger efficiency.

On the other hand, the inclusive rate was measured independently from the bunch-by-bunch rate measurement.
As a cross-check, the inclusive rate measurement with the same method as Scan-III using Eq.\ref{eq:inclusive-cs}
was performed, giving 3.92$\pm$0.03~barn with a simple Gaussian + straight line baseline fit,
and 3.97$\pm$0.04~barn with numerical shape analysis method using Eq.\ref{eq:cross}.
In both cases, there is no additional correction applied.
These results are $\sim$ 2\% different from those obtained by using the bunch-by-bunch analysis method.
This latter is the most accurate method and has been adopted as the central value.

\subsection{Systematic Uncertainty}

\begin{table}[tb]
   \centering
   \caption{Relative systematic uncertainties for \PbPb\ scans.}
   \vspace{2mm}
   \begin{tabular}{llll}
       \toprule
       \textbf{items}              & \textbf{Scan-III}     & \textbf{Scan-VIII} \\
       \midrule
       bunch intensity             &                       &                    \\
       \midrule
       DCCT scale                  & 2.7\%                 & 0.4\%              \\
       relative bunch populations  & $<$0.1\%              & $<$0.1\% \\
       ghost charge                & $_{-1.4}^{+3.9}\:$\%  & $_{-1.1}^{+4.4}\:$\% \\  
       satellites                  & 0.5\%                 & 0.5\% \\
       \midrule
       experiment                  &                       &                    \\
       \midrule
       length scale calibration    & 2$\oplus$2\%          & 1$\oplus$1\%  \\
       luminosity decay            & 2\%                   & 2\% \\
       inclusive v.s. b-by-b       & 2\%                   & N.A. \\
       background                  & 1\%                   & 1\%  \\
       scan-to-scan discrepancy    & 1\%                   & 1\%  \\
       \midrule
       total                       &  $_{-5.2}^{+6.4}\:$\% & $_{-3.1}^{+5.3}\:$\% \\
       \bottomrule
   \end{tabular}
   \label{table-pbpb-uncert}
\end{table}

Table~\ref{table-pbpb-uncert} shows the summary of systematic uncertainties considered for the ZED cross section
for Scan-III and VLN cross section for Scan-VIII.

The bunch intensity values were given by BCNWG\cite{BCNWG-DCCT}.
The largest uncertainties in the bunch intensity measurements are the DCCT calibration uncertainty
(2.7\%) and the uncertainty in ghost charge correction ($_{-1.4}^{+3.9}$\%)\cite{SATELLITE} for Scan-III.
It was improved in Scan-VIII to 0.4\% of DCCT scale uncertainty and $_{-1.1}^{+4.4}$\% for ghost charge correction
uncertainty.
The ghost charge uncertainty shown in the Table~\ref{table-pbpb-uncert} is translated to the
uncertainty on the cross section via $(1-g_1)\cdot(1-g_2)$ where $g_1$ and $g_2$ are measured fractions of ghost charge
in Beam-1 and Beam-2, knowing that the uncertainties on $g_1$ and $g_2$ are fully correlated.

The length scale calibration uncertainty is the same as for the \pp\ cases.
The LSC data is missing for the Scan-III, so we assign a systematic uncertainty of
2\% per direction, corresponding to a conservative estimate of the maximum correction factor.
Scan-VIII has LSC, thus a smaller uncertainty is assigned.

Since lead beam losses were much higher than that of proton beams,
the luminosity decay correction was larger.
After it is applied, as shown in Table~\ref{table-pbpb}, $R^\mathrm{top}$ still shows a 2\% discrepancy
which indicates residual emittance growth or other effects due to changes of beam properties.
This was assigned as a systematic uncertainty.

The method of cross section calculation by Eq.~\ref{eq:inclusive-cs} is not perfectly accurate because
the beam parameters are different among all bunch pairs.
The accurate method is given by Eq.\ref{eq:bbb-cs}, averaging individually measured cross sections.
Because the bunch-by-bunch measurement of rates was not possible in Scan-III,
neither $\sigma_{x,y}^i$ nor $R_{i}^\mathrm{top}$ are available.
Therefore, the method given by Eq.~\ref{eq:inclusive-cs} introduces systematic uncertainty.
However the bunch-by-bunch measurement became available from Scan-V onward.
Therefore, the knowledge obtained in Scan-VIII was used to estimate such a systematic uncertainty.
As described in the previous subsection,
$\SVLN$ values were measured by both accurate bunch-by-bunch analysis and inclusive analysis,
where the discrepancy was found to be 2\% and this was assigned as the systematic uncertainty in Scan-III result.
Additionally, a simulation with a realistic variation of beam sizes and intensity showed a less than
2\% discrepancy and, hence, 
supports that 2\% is conservative enough at this moment.
Scan-VIII does not have such systematic uncertainty.

The background level, producing a  baseline in rate measurements, may have as much as 1\% effect on the cross section value.
This was estimated from the rate plot shown in Fig.~\ref{fig-shapePbPb}.
The baseline was fixed at about 190~Hz.
Although varying it by 10\% to 20\% which is obviously out of the realistic range,
the effect on cross section was found to be below 1\%
because results are dominated by the trigger rate data at a smaller separation where the rate is higher.
Thus 1\% as the systematic uncertainty is still conservative.

Since a 0.9\% discrepancy is observed between two sets of scans, 1\% was assigned as the scan-to-scan discrepancy.

Thus the total relative uncertainty for $\SZED$ is estimated as
$_{-5.2}^{+6.4}\:$\%, and for $\SVLN$ as $_{-3.1}^{+5.3}\:$\%.
Statistical uncertainties are negligibly small.

\section{Conclusions}\label{Sec:Conclusion}

Several reference cross sections were measured in the ALICE experiment using the van der Meer scan technique.

The reference cross section ($\SV$) of the ALICE VZERO detector
has been measured for 7 TeV and 2.76 TeV \pp\ collisions to be
$\SV(\mbox{7 TeV})    =  54.34 \pm 1.90\mbox{(syst.)} \; \mbox{mb}$ and
$\SV(\mbox{2.76 TeV}) =  47.67 \pm 0.88\mbox{(syst.)} \; \mbox{mb}$.
The results were used for cross section determinations of other physics processes in ALICE.

The reference cross section of the ALICE neutron ZDC trigger ($\SZED$), sensitive for electromagnetic
nuclear dissociation process, has been obtained for $\sqrtsNN=2.76$~TeV \PbPb\ collisions as
$371.4_{-19.3}^{+23.8}$(syst.)$\pm 0.6$(stat.)~barn.

The cross section $\SVLN$ for VLN logic sensitive to $\simeq$50$\%$ most central hadronic Pb-Pb collisions
has been calculated as 4.10~$^{+0.22}_{-0.13}$(syst.)~barn
in reasonable agreement with expected cross section value, considering the nuclei size and the trigger efficiency.

\end{document}